\documentclass[twocolumn]{aa}
\usepackage[utf8]{inputenc}
\usepackage{esint}
\usepackage{verbatim}
\usepackage{blkarray,bigstrut}
\usepackage{mathtools}
\usepackage{cuted}
\usepackage{amsmath}
\usepackage{bm}
\usepackage{widetext}
\usepackage{esvect}
\usepackage{hyperref}
\usepackage{xcolor}
\usepackage{placeins}
\usepackage{xspace}
\usepackage[ruled,vlined,linesnumbered]{algorithm2e}
\newcommand{\numBHB}{90}
\newcommand{\GWT}{\texttt{GW-Universe Toolbox}\xspace}

\begin{document}
    \title{The GW-Universe Toolbox II: constraining the binary black hole population with second and third generation detectors}
    \author{Shu-Xu Yi\inst{1,5}\thanks{sxyi@ihep.ac.cn}, Fiorenzo Stoppa\inst{1,2}, Gijs Nelemans\inst{1,3,4}, Eric Cator\inst{2}}
    \institute{Department of Astrophysics/IMAPP, Radboud University,
    P.O. Box 9010, NL-6500 GL Nijmegen, The Netherlands\\
    \and
    Department of Mathematics/IMAPP, Radboud University, 
    P.O. Box 9010, NL-6500 GL Nijmegen, The Netherlands\\
    \and 
          SRON, Netherlands Institute for Space Research, Sorbonnelaan 2, NL-3584 CA Utrecht, The Netherlands
         \and
         Institute of Astronomy, KU Leuven, Celestijnenlaan 200D, B-3001 Leuven, Belgium
    \and
    Key Laboratory of Particle Astrophysics, Institute of High Energy Physics, Chinese Academy of Sciences, 19B Yuquan Road, Beijing 100049, People’s Republic of China}

    \date{\today}
    \abstract{
    {\it Context: } The \GWT is a software package that simulates observations of the Gravitational Wave (GW) Universe with different types of GW detectors, including Earth-based/Space-borne laser interferometers and Pulsar timing arrays. It is accessible as a website, and can also be imported and run locally as a Python package.\\
    {\it Methods:}
    We employ the method used by the \GWT to generate a synthetic catalogue of detection of stellar mass binary black hole (BBH) mergers. As an example of its scientific application, we study how GW observations of BBHs can be used to constrain the merger rate as function of redshift and masses. We study advanced LIGO (aLIGO) and Einstein Telescope (ET) as \emph{two representatives for the 2nd and 3rd generation GW observatories}. We also simulate 
    the observations from a detector that is half as sensitive as the ET at design which represents an early phase of ET. Two methods are used to obtain the constraints on the source population properties from the catalogues: the first uses a parameteric differential merger rate model and applies a Bayesian inference on the parameters; The other one is non-parameteric and uses weighted Kernel density estimators.\\
    {\it Results:} The results show the overwhelming advantages of the 3rd generation detector over the 2nd generation for the study of BBH population properties, especially at a redshifts higher than $\sim2$, where the merger rate is believed to peak. With the simulated aLIGO catalogue, the parameteric Bayesian method can still give some constraints on the merger rate density and mass function beyond its detecting horizon, while the non-parametric method lose the constraining ability completely there. The difference is due to the extra information placed by assuming a specific parameterisation of the population model in the Bayesian method. In the non-parameteric method, no assumption of the general shape of the merger rate density and mass function are placed, not even the assumption of its smoothness. These two methods represent the two extreme situations of general population reconstruction. We also find that, despite the numbers of detection of the half-ET can be easily compatible with full ET after a longer observation duration, the catalogue from the full ET can still give much better constraints on the population properties, due to its smaller uncertainties on the physical parameters of the GW events. }
   
\maketitle

\section{Introduction}
The era of gravitational wave (GW) astronomy has begun with the first detection of a binary black hole (BBH) merger in 2015 \citep{2016PhRvL.116f1102A}. There have been $\sim\numBHB$ BBH systems discovered with $\sim3$ years observation in the LIGO/Virgo O1/O2/O3 runs (GWTC-1\&2 catalogues: \cite{2019PhRvX...9c1040A, 2020arXiv201014527A}). The number is catching up on that of the discovered Galactic black holes (BH) in X-ray binaries during the last half century \citep{2016A&A...587A..61C}. LIGO and Virgo are being upgraded towards their designed sensitivity, and new GW observatories such as The Kamioka Gravitational Wave Detector (KAGRA, \citealt{2019NatAs...3...35K}) and LIGO-India \citep{indigo} will join the network in the near future. Furthermore, there are plans for the next generation of GW observatories such as the Einstein Telescope (ET, \citealt{2010CQGra..27s4002P}) and the Cosmic Explore (CE, \citep{2019BAAS...51g..35R}), which will push forward the detecting horizon significantly. With the expectation of the exploding size of the BBH population, the community has been actively studying what science can be extracted from GW observations of BBHs. Applications include e.g. the cosmic merger rate density of BBH \citep[e.g.][]{2019ApJ...886L...1V}, the mass distribution of stellar mass black hole \citep[e.g.][]{2017PhRvL.119m1301K,2017PhRvD..95j3010K,2019ApJ...882L..24A,2021ApJ...913L...7A}, the rate of stellar nucleosynthesis \citep[e.g.][]{2020arXiv200606678F} and constraints on the expansion history of the universe \citep{2019ApJ...883L..42F}. Simulating a catalogue of observations is the crucial step in the above mentioned work. In all the previous literature, the simulation is performed for specific detectors and observation duration and their catalogues are difficult to extend to other usages. Therefore, we are building a set of simulating tools which has the flexibility to different detectors, observation duration, source populations and cosmological models. The \GWT is such a tool set that simulates observations of the GW Universe (including populations of stellar mass compact binary mergers, supermassive black hole binaries inspiral and mergers, extreme mass ratio inspirals, etc.) with different types of GW detectors including Earth-based/Space-borne laser interferometers and Pulsar timing arrays. It is accessible as a website\footnote{\url{gw-universe.org}}, or run locally as a Python package. For a detailed overview of the functionalities and methods of \GWT, please see \citealt{2021arXiv210613662Y}, referred to as the \textit{Paper-I} hereafter.

Here we exhibit one application of the \GWT by showing how the cosmic merger rate and mass function of BH can be constrained by GW observations using advanced LIGO (aLIGO, at design sensitivity) and ET \footnote{The current \GWT can only simulate observations with single detectors. A network of detectors is expected to be more efficient in the following aspects: 1. the total SNR of a joint-observed GW source would be larger (but in the same order of magnitude) than that from a single detector; 2. the uncertainties of source parameters would be smaller. In the current \GWT, the duty cycle of detectors are assumed to be 100\%. In a more realized simulation, where the duty cycles of single detectors are not 100\%, a detector network also benefits from a larger time coverage. }. The cosmic merger rate density, or the volumetric merger rate as function of redshift $\mathcal{R}(z)$, provides valuable information on the history of the star formation rate (SFR), which is crucial for understanding the evolution of the Universe, e.g, the chemical enrichment history, and the formation and evolution of galaxies \citep{2014ARA&A..52..415M}. The SFR as function of redshift has been studied for decades mainly with UV/optical/infrared observation of galaxies in different redshift \citep[][]{2002MNRAS.330..621S,2006MNRAS.370..331D,2006ApJ...647..787T,2008ApJ...689..883C,2010ApJ...714.1740M,2010ApJ...718.1171R,2011ApJ...726..109L,2011MNRAS.416.1862S,2012ARA&A..50..531K,2013MNRAS.429..302C,2014ARA&A..52..415M}. Such electromagnetic (EM) wave probes are limited by dust extinction and the completeness of samples at high redshifts is difficult to estimate \citep[e.g.][]{2019A&ARv..27....3M}. Moreover, the connection between the direct observable and the SFR history is model dependent on multiple layers \citep[see e.g.][]{2021MNRAS.508.4994C}. GW provide a unique new probe which suffer less from the above mentioned limitations. 

Moreover, $\mathcal{R}(z)$ depends also on the distribution of delay times $\mathbf{\tau}$, i.e., the time lag between the formation of the binary massive stars and the merger of the two compact objects \cite[e.g.][]{2010ApJ...715L.138B,2018MNRAS.481.1908K}. Therefore, a combination of the information of the SFR from EM observations and $\mathcal{R}(z)$ from GW can improve our knowledge of the delay time distribution. $P(\tau)$ together with primary mass function $p(m_1)$ (where $m_1$ is the mass of the primary BH) of BBH helps us to better understand the physics in massive binary evolution, e.g., in the common envelope scenario \cite[e.g.][]{2016Natur.534..512B,2018MNRAS.481.1908K}. 

There are several previous works that studied the prospect of using simulated GW observations to study $\mathcal{R}(z)$ and $p(m_1)$ of BBH. \cite{2019ApJ...886L...1V} shows that with three months observation with a third-generation detector network, $\mathcal{R}(z)$ can be constrained to high precision up to $z\sim 10$. In their work the authors did not take into account the signal-to-noise (SNR) limits of detectors and instead made the assumption that all BBH mergers in the Universe are in the detected catalogue. \cite{2017PhRvD..95j3010K} studied how the observation with aLIGO can be used to give a constraint on $p(m_1)$. They assumed a certain parameterization of $p(m_1)$, and use an approximated method to obtain the covariance of the parameters based on a simulated catalogue of aLIGO observations. Recently, \cite{2021arXiv211204058S} simulated observation with ET on compact object binary mergers, taking into account the effect of the rotation of the Earth. They found that $\mathcal{R}(z)$ and mass distribution can be so well constrained from ET's observation, that mergers from different populations can be distinguished. 

In this work, we simulate the observed catalogues of BBH mergers in ten years of aLIGO observation with its design sensitivity and one month ET observation using the \GWT. We also simulate 
    the observation from a detector that is half as sensitive as the ET in design which represents the early phase of ET. The paper is organised as follows: In section \ref{sec:thoery}, we summarize the method used by the \GWT to simulate the catalogue of BBH mergers; In section \ref{sec:constraint}, we use the synthetic catalogues to set constraints on the cosmic merger rate and mass function of the population. We employ two ways to constrain $\mathcal{R}(z)$ along with the mass function of the primary BH. The first assumes parameteric forms of $\mathcal{R}(z)$ and the mass function and uses Bayesian inference on their parameters. The other method does not assume a parameteric formula and uses weighted Kernel density estimators. The results are presented and compared. We conclude and discuss in section \ref{sec:candd}.  

\section{Generating synthetic catalogues}\label{sec:thoery}
\subsection{method}
\begin{figure}
    \centering
    \includegraphics[width=0.5\textwidth]{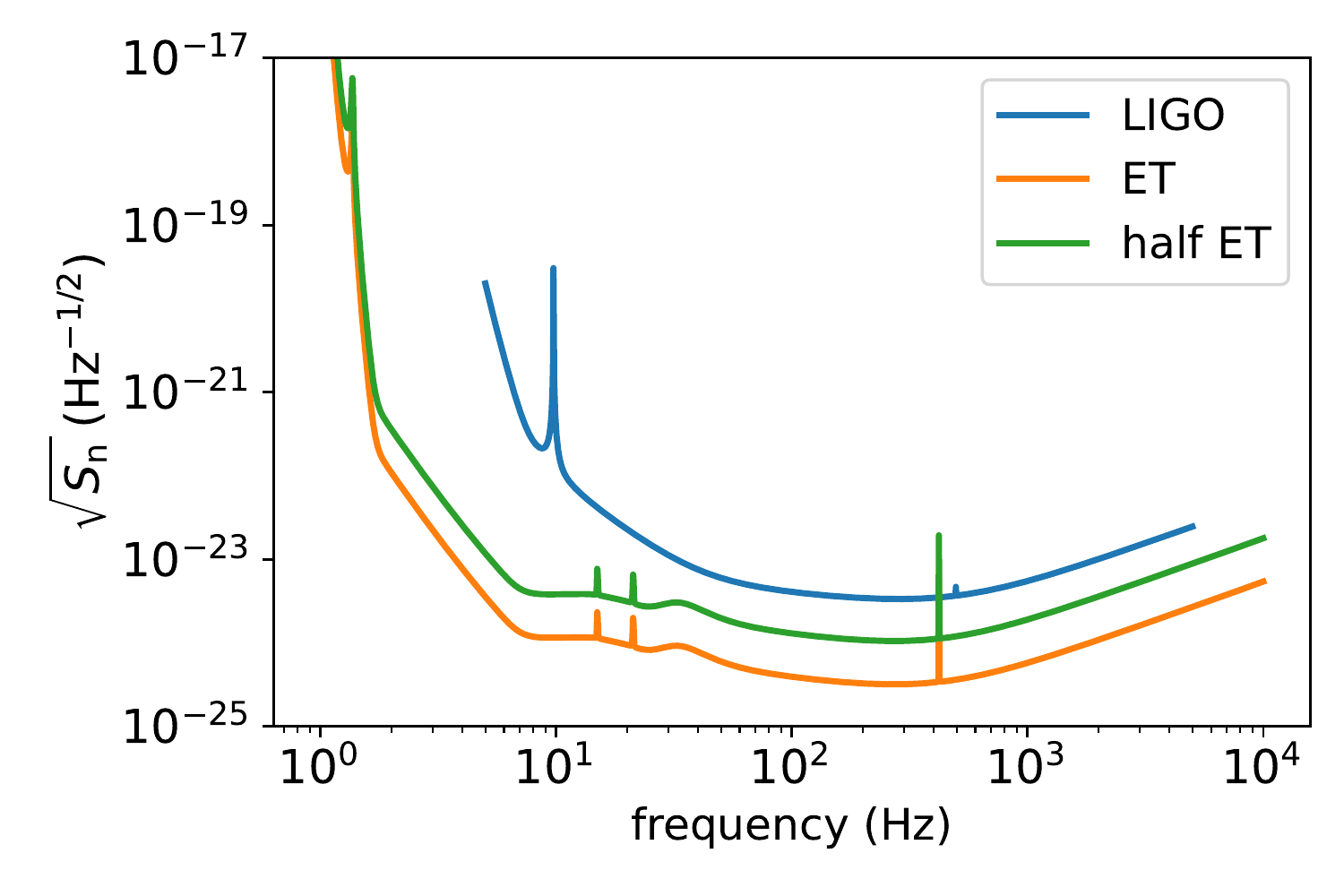}
    \caption{The squared-root of the noise power spectrum density of the aLIGO (blue) and ET (orange) at their designed sensitivity. The data of the plots are obtained from \url{https://dcc.ligo.org/LIGO-T1800044/public} and \url{http://www.et-gw.eu/index.php/etsensitivities}}
    \label{fig:Sn}
\end{figure}

We use the \GWT to simulate observations on BBH with three different detectors, namely, aLIGO (designed sensitivity), ET and ``half-ET" \footnote{Here we are referring to the ET-D design. For the key parameters of this ET configuration, see \cite{2011CQGra..28i4013H}.}. The noise power spectrum $S_{\rm{n}}$ for aLIGO and ET are obtained from \footnote{\url{https://dcc.ligo.org/LIGO-T1800044/public} \\ \url{http://www.et-gw.eu/index.php/etsensitivities}}. The ``half-ET" represents our expectation on the performance of the ET in its early phase. 
In figure \ref{fig:Sn}, we plot the $\sqrt{S_{\rm{n}}}$ as function of frequency for aLIGO, ET at their designed sensitivity, along with that of the half-ET. For its $S_{\rm{n}}$, we multiply $S_{\rm{n}}$ of the ET with a factor 10.89 to obtain a noise curve with is in the half way between the designed aLIGO and ET.
Here we will give a brief summary of the method. For details, we refer to \cite{2021arXiv210613662Y}. We first calculate a function $\mathcal{D}(z, m_1, m_2)$, which represents the detectability of a BBH with redshift $z$, primary and secondary masses $m_1$, $m_2$. $\mathcal{D}(z, m_1, m_2)$ depends on the antenna pattern and $S_{\rm{n}}$ of the detector, and the user designated signal-to-noise (SNR) threshold. The other ingredient is the merger rate density distribution of BBH in the Universe, $\dot{n}(z, m_1, m_2)$. In the \GWT, we integrate two parameterized function forms for $\dot{n}(z, m_1, m_2)$, namely pop-A and pop-B (see \citealt{2021arXiv210613662Y} for the description on the population models). For more sophisticated $\dot{n}(z, m_1, m_2)$ from population synthesis simulations, see \cite{2002ApJ...572..407B,2017MNRAS.472.2422M,2018MNRAS.479.4391M,2021arXiv210912119A,2021arXiv210804250B,2021arXiv210906222M,2021arXiv211001634V,2021arXiv211204058S}. 

The distribution of detectable sources, which we denote as $N(z,m_1,m_2)$ can be calculated with: 
\begin{equation}
    N_{\rm{D}}(\mathbf{\Theta})=\frac{T}{1+z}\frac{dV_{\rm{c}}}{dz}\dot{n}(\mathbf{\Theta})\mathcal{D}(\mathbf{\Theta}),\label{eqn:detectables}
\end{equation}
where $T$ is the observation duration, and $\mathbf{\Theta}=(z,m_1,m_2)$. An  integration of $N_{\rm{D}}(\mathbf{\Theta})$ gives the expectation of total detection number:
\begin{equation}
    N_{\rm{tot,exp}}=\int d\mathbf{\Theta}N_D(\mathbf{\Theta}).
\end{equation}
The actual total number in a realization of simulated catalogue is a Poisson random around this expectation. The synthetic catalogue of GW events is drawn from $N_D(\mathbf{\Theta})$ with a Markov-Chain Monte-Carlo (MCMC) algorithm. 

We can also give a rough estimation of the uncertainties on the intrinsic parameters of each event in the synthetic catalogue using the Fisher Information Matrix (FIM). 
Previous studies ({\it e.g.,} \cite{2013PhRvD..88h4013R}) found that the Fisher information matrix will sometimes severely overestimated the uncertainties, especially when the SNR is close to the threshold. We implement a correction that if the relative uncertainty of masses calculated with FIM is larger than $20\%$, $\delta m_{1,2}=0.2m_{1,2}$ is applied instead. 

The uncertainty of the redshift is in correlation with those of other external parameters, such as the localization error. The actual localization uncertainties are largely determined by the triangulation error of the detector network, which can not be estimated with the current version of \GWT\  yet. As a result, we do not have a reliable way that accurately estimates the errors on redshift. As a {\it ad hoc} method, we assign $\delta z=0.5z$ as a typical representative for the uncertainties of aLIGO, and $\delta z=0.017z+0.012$ as a typical representative for the uncertainties of ET \citep{2017PhRvD..95f4052V}. For half-ET, we assign $\delta z=0.1z+0.01$. 

\subsection{Catalogues}\label{cat}

\begin{figure}
    \centering
    \includegraphics[width=0.5\textwidth]{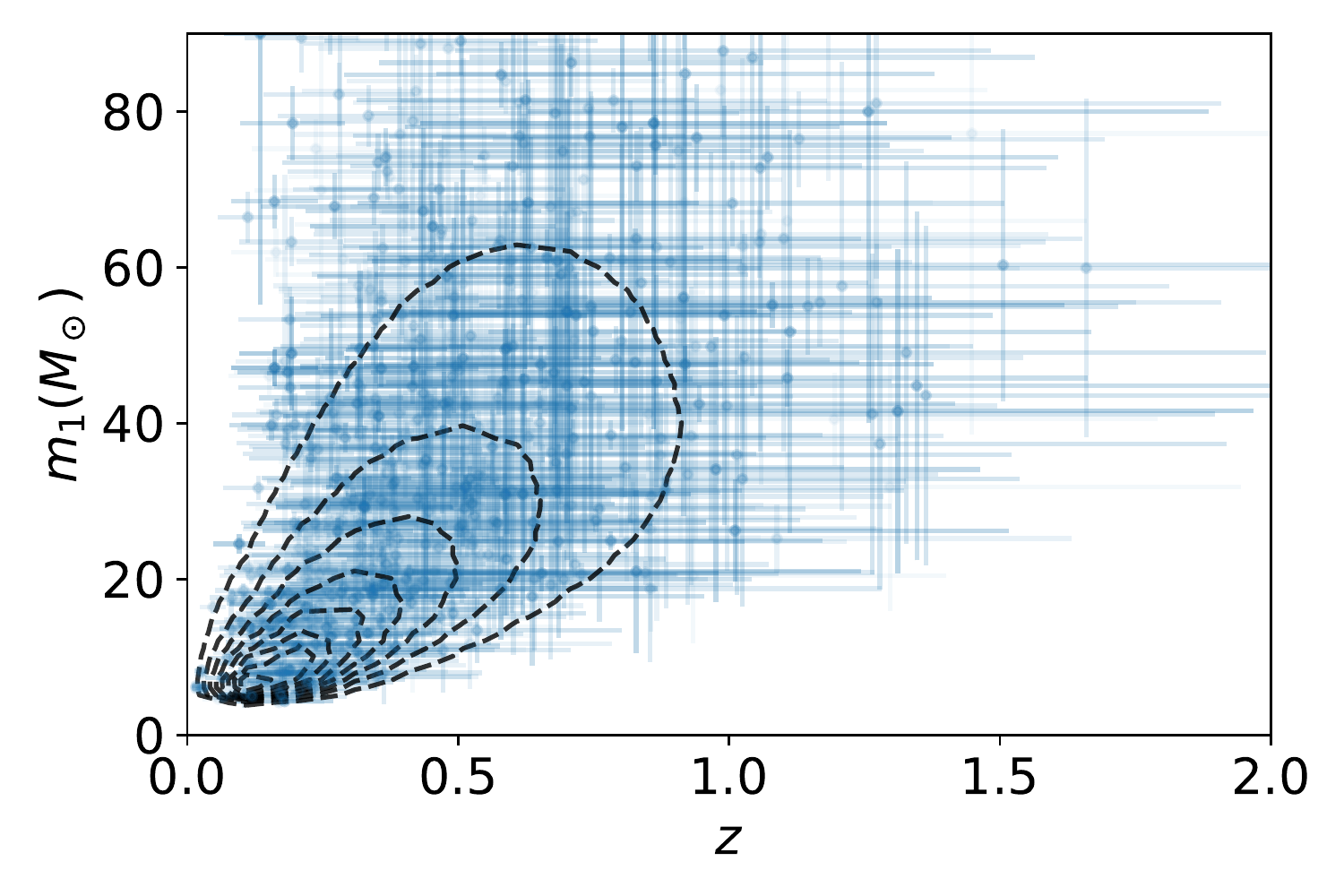}
    \caption{\textbf{aLIGO 10 years' catalogue: } Black dashed lines are contours of $N_D(\mathbf{\Theta})$; Blue points with error bars are 2072 events in the simulated catalogues.}
    \label{fig:figure2}
\end{figure}

\begin{figure}
    \centering
    \includegraphics[width=0.5\textwidth]{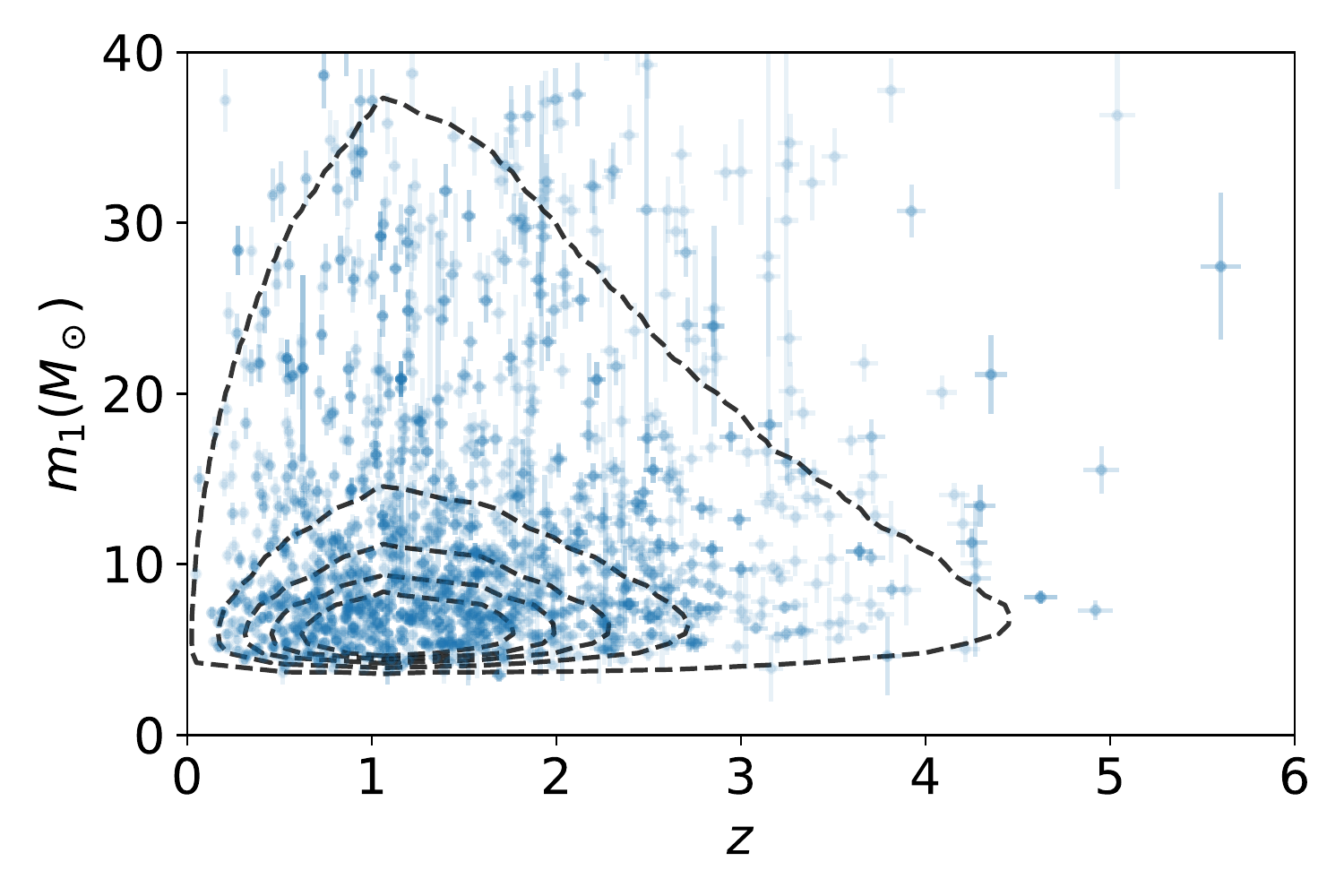}
    \caption{\textbf{ET 1 month's catalogue: } Black dashed lines are contours of $N_D(\mathbf{\Theta})$; Blue points with error bars are 1830 events in the simulated catalogues.}
    \label{fig:figure3}
\end{figure}

\begin{figure}
    \centering
    \includegraphics[width=0.5\textwidth]{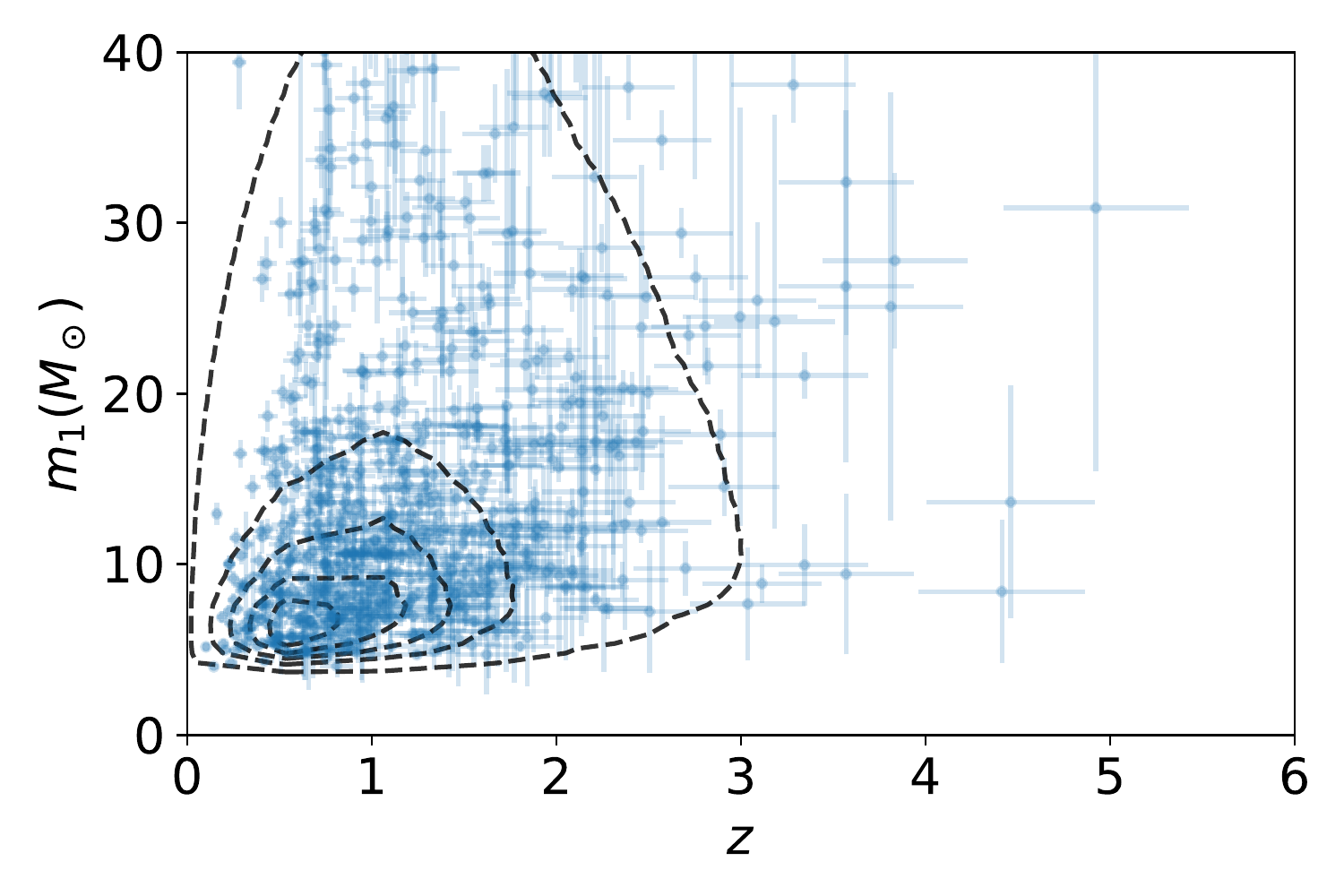}
    \caption{\textbf{half-ET 1 month's catalogue: }Black dashed lines are contours of $N_D(\mathbf{\Theta})$; Blue points with error bars are 889 events in the simulated catalogues.}
    \label{fig:figure4}
\end{figure}

Using the above-mentioned method, we generate synthetic catalogues corresponding to 10-years aLIGO, 1-month ET observations and 1-month observation with half-ET. In figures \ref{fig:figure2}, \ref{fig:figure3} and \ref{fig:figure4}, we plot events in those catalogues, along with the corresponding marginalized $N_D(\mathbf{\Theta})$. The numbers in the catalogues are 2072, 1830 and 889 for the 10-year aLIGO, 1-month ET and 1-month half-ET respectively. The catalogues agree with the theoretical number density, which proves the validity of the MCMC sampling process. 

The underlying population model for $\dot{n}(z,m_1,m_2)$ is pop-A, with parameters: $R_{\rm{n}}=13$\,Gpc$^{-3}$yr$^{-1}$, $\tau=3$\,Gyr, $\mu=3$, $c=6$, $\gamma=2.5$, $m_{\rm cut}=95\,M_\odot$, $q_{\rm cut}=0.4$, $\sigma=0.1$. Those parameters are chosen because they result in simulated catalogues which are in compatible with the real observations (see \citealt{2021arXiv210613662Y}). 

\section{Constraints on $\mathcal{R}(z)$ and $p(m_1)$ from BBH catalogues}\label{sec:constraint}
In this section, we study how well $\mathcal{R}$ and $p(m_1)$ can be constrained with BBH catalogues from observations with 2nd and 3rd generations GW detectors. Two distinct methods will be used. 


\subsection{Bayesian method} 
In the first method, we use a parameterised form of $\dot{n}(\mathbf{\Theta}|\mathcal{B})$. $\mathcal{B}$ are the parameters describing the population rate density, which are $\mathcal{B}=(R_0,\tau,\mu,c,\gamma,q_{\rm cut},\sigma)$ in our case. $\mathbf{\Theta}^k$ are the physical parameters of the $k$-th source in the catalogue, and we use $\{\mathbf{\Theta}\}$ to denote the whole catalogue. 

The posterior probability of $\mathcal{B}$ given an observed catalogue $\{\mathbf{\Theta}\}_{\rm obs}$:
\begin{equation}
    p(\mathcal{B}|\{\mathbf{\Theta}\}_{\rm obs})\propto p(\{\mathbf{\Theta}\}_{\rm obs}|\mathcal{B}) p(\mathcal{B}),
    \label{eqn:bayesian}
\end{equation}
where $p(\{\mathbf{\Theta}\}_{\rm obs}|\mathcal{B})$ is the likelihood and $p(\mathcal{B})$ is the prior. The subscript ``obs" is to distinguish the observed parameters from the true parameters of the sources. The likelihood of detecting a catalogue $\{\mathbf{\Theta}\}_{\rm obs}$ given the parameters $\mathcal{B}$ is a known result as an inhomogeneous Poisson Process \citep{2014ApJ...795...64F}:
\begin{equation}
    p(\{\mathbf{\Theta}\}_{\rm obs}|\mathcal{B})=N!\prod_{k=1}^{N}p(\mathbf{\Theta}^k_{\rm obs}|\mathcal{B})p(N|\lambda(\mathcal{B})),
    \label{eqn:lkhood}
\end{equation}
where $p(N|\lambda(\mathcal{B}))$ is the probability of detecting $N$ sources if the expected total number is $\lambda$, which is a Poisson function. The probability of detecting a source $k$ with parameters $\mathbf{\Theta}^k$ is:
\begin{equation}
    p(\mathbf{\Theta}^k_{\rm obs}|\mathcal{B})=\int p(\mathbf{\Theta}^k_{\rm obs}|\mathbf{\Theta}^k_{\rm true})p(\mathbf{\Theta}^k_{\rm true}|\mathcal{B})d\mathbf{\Theta}^k_{\rm true}.
    \label{eqn:integration}
\end{equation}
We further assume that observational errors are Gaussian, therefore we have $p(\mathbf{\Theta}^k_{\rm obs}|\mathbf{\Theta}^k_{\rm true})=p(\mathbf{\Theta}^k_{\rm true}|\mathbf{\Theta}^k_{\rm obs})$. Taking this symmetric relation into equation \ref{eqn:integration}, we have:
\begin{equation}
    p(\mathbf{\Theta}^k_{\rm obs}|\mathcal{B})=\int p(\mathbf{\Theta}^k_{\rm true}|\mathbf{\Theta}^k_{\rm obs})p(\mathbf{\Theta}^k_{\rm true}|\mathcal{B})d\mathbf{\Theta}^k_{\rm true}.
    \label{eqn:integration_2}
\end{equation}
The above integration is equivalent to the average of $p(\mathbf{\Theta}^k_{\rm true}|\mathcal{B})$ over all possible $\Theta^k$. Therefore:
\begin{equation}
    p(\mathbf{\Theta}_{\rm obs}|\mathcal{B})\approx\left<p(\mathbf{\Theta}^k_{\rm true}|\mathcal{B})\right>,
    \label{eqn:MCintegrate}
\end{equation}
where the $\left<\cdots\right>$ denotes an average among a sample of $\mathbf{\Theta}^k$ which is drawn from a multivariate Gaussian characterized by the observation uncertainties. Taking equation (\ref{eqn:MCintegrate}) into equation (\ref{eqn:lkhood}), and noting that $p(\mathbf{\Theta}^k|\mathcal{B})\equiv N_D(\mathbf{\Theta}^k|\mathcal{B})/\lambda(\mathcal{B})$, and $p(N|\lambda(\mathcal{B}))$ is the Poisson distribution:
\begin{equation}
    p(\{\mathbf{\Theta}\}_{\rm obs}|\mathcal{B})=\prod_{k=1}^{N}\left<N_D(\mathbf{\Theta}^k|\mathcal{B})\right>\exp(-\lambda(\mathcal{B})).
\end{equation}
For the prior distribution $p(\mathcal{B})$, we use the log normal distributions for $\tau$, $R_0$, $c$, $\gamma$ and $\mu$, which center at the true values and large enough standard deviations. The posterior probability distributions of the parameters are obtained with MCMC sampling from the posterior in equation (\ref{eqn:bayesian}). We keep the nuisance parameters $q_{\rm{cut}}$ and $\sigma$ fixed to the true value, in order to reduce the necessary length of chain. The posterior of the interesting parameters would not be influenced significantly by leaving these free. 
\begin{figure}
    \centering
    \includegraphics[width=0.5\textwidth]{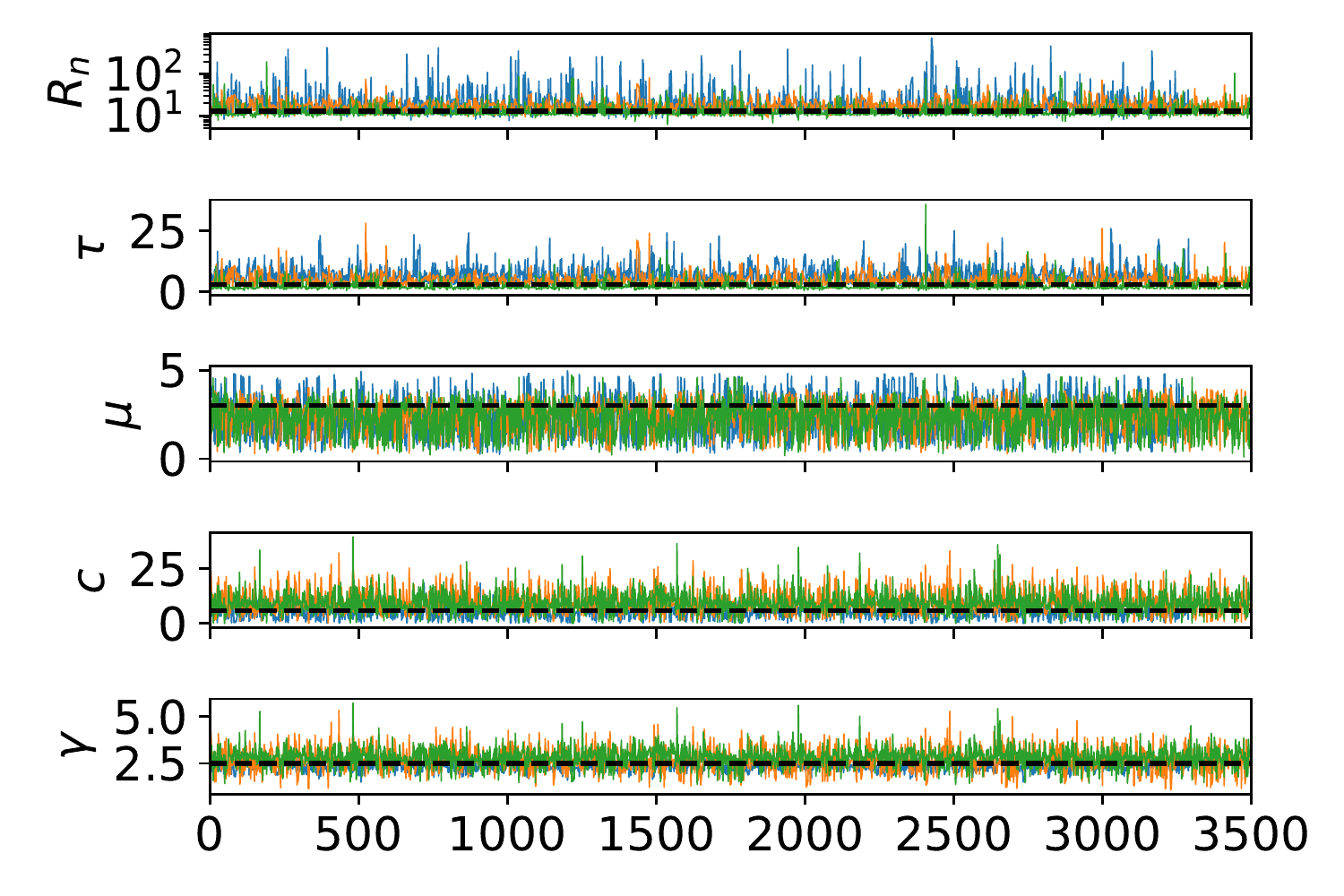}
    \caption{The MCMC chains sampled from the posterior distribution of the population parameters. \textbf{The blue line} is for aLIGO-10 years observation; \textbf{The green line} is for half-ET's 1 month observation, and \textbf{the orange line} is for full ET's 1 month observation. The black dashed lines mark the true values of the corresponding parameters, which were used to generate the observed catalogue.}
    \label{fig:chain}
\end{figure}
In Figure \ref{fig:chain}, we plot the MCMC chains sampled from the posterior distribution of the population parameters, where we can see the chains are well mixed, and thus can serve as good representation of the underlying posterior distribution. In Figure \ref{fig:cor}, we plot the correlation between the parameters of the red-shift dependence ($R_{\rm{n}}$, $\tau$) and among the parameters of the mass function ($\mu$, $c$, $\gamma$). As can be found, there are strong correlations between the $\mathcal{R}(z)$ parameters $\tau$ and $R_{\rm{n}}$, and also among the $p(m_1)$ parameters $\gamma$, $c$ and $\mu$. The correlation between parameters of the red-shift dependence and those of the mass function are small and therefore not plotted. 

\begin{figure*}[!h]
    \centering
    \includegraphics[width=\textwidth]{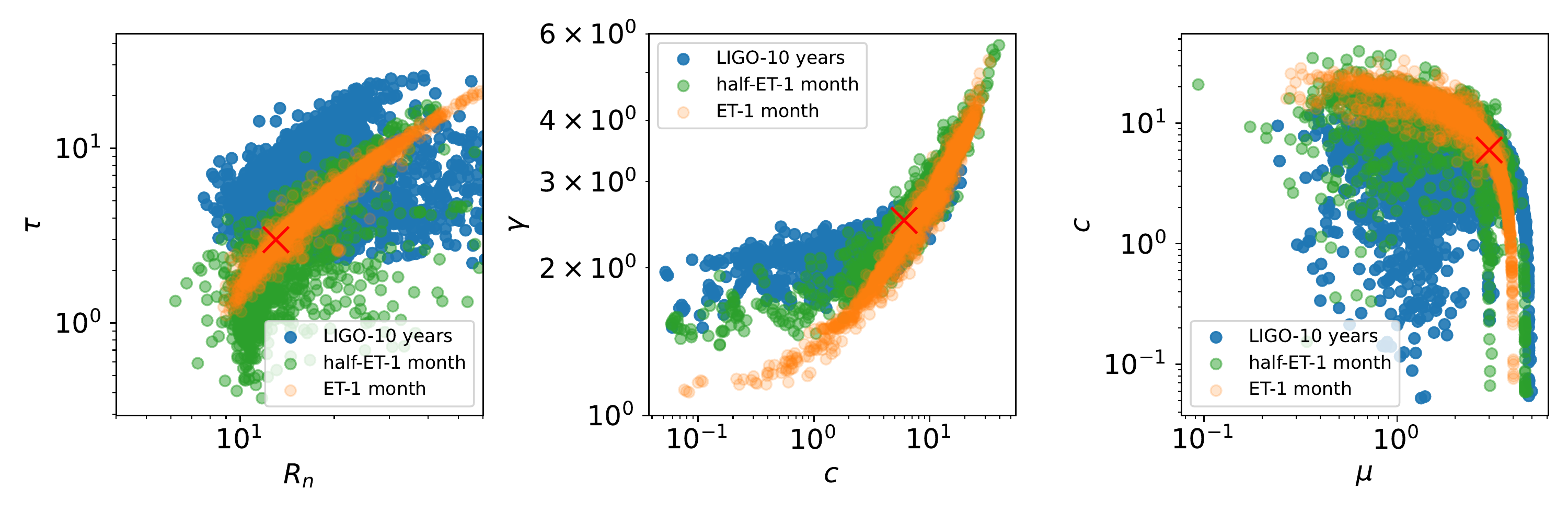}
    \caption{The correlation between the parameters of red-shift dependence ($R_{\rm{n}}$, $\tau$, left panel) and among the parameters of mass function ($\mu$, $c$, $\gamma$, middle and right panel). Blue markers correspond to 10 years observation with aLIGO; Green markers correspond to 1 month observation with half-ET, and orange markers correspond to 1 month observation with ET. The red cross in each panel marks the location of the true values, which were used to generate the observed catalogue.}
    \label{fig:cor}
\end{figure*}
\begin{figure}
    \centering
    \includegraphics[width=0.5\textwidth]{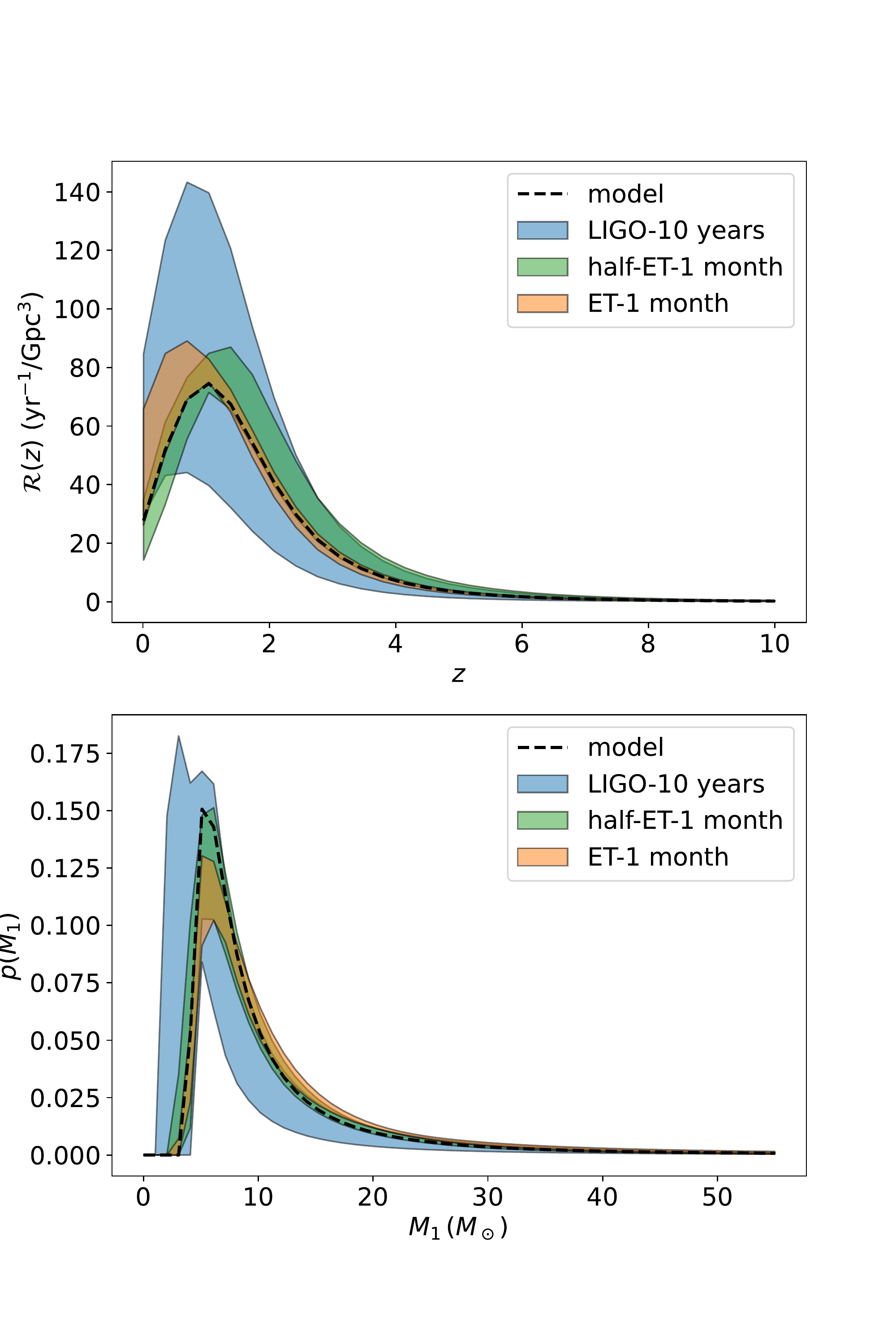}
    \caption{$\mathcal{R}(z)$ (upper panel) and $p(m_1)$ (lower panel) reconstructed with the Bayesian method, from the synthetic aLIGO (blue), ET (orange) and half-ET (green) catalogues. The black dashed curves are the model used to generate the catalogues.}
    \label{fig:reconstruct}
\end{figure}

In figure \ref{fig:reconstruct}, we plot the inferred differential merger rate as function of redshift and mass of the primary BH. The bands correspond to the 95\% confidence level of the posterior distribution of the population parameters. We can see from the panels of figure \ref{fig:reconstruct} that, although ten years of aLIGO observation and 1-month of ET observation result in a similar number of detections, ET can still obtain much better constraints on both $\mathcal{R}(z)$ and $p(M_1)$. That results from 1): the much smaller uncertainties on parameters with ET observation compared to aLIGO, and 2): The catalogue observed by ET covers larger redshift and mass ranges than that of aLIGO. For half-ET and ET, at low red-shift and high mass, both detectors result in a similar number of detections. Therefore, they give comparable  constraints on $\mathcal{R}(z)$ at low redshift and $p(m_1)$ at large mass. While at high red-shift and low mass, the full ET observed more events and with lower uncertainties, therefore gives better constrains than that of half-ET. 

We see from figure \ref{fig:figure2} that the maximum red-shift in aLIGO's catalogue is $z\sim1.5$, while $\mathcal{R}(z)$ can still be constrained for $z>1.5$, and even be more stringent towards larger red-shift. This constraint on $\mathcal{R}(z)$ at large redshift is not from the observed catalogue, but is intrinsically imposed by the assumed model in the Bayesian method: we have fixed our star formation rate as function of redshift so that $\psi(z_f)$  peaks at $z\sim2$, the peak of BBH merger rate can therefore only appear later. 

\subsection{Nonparametric method}
From equation (\ref{eqn:detectables}), we know that:
\begin{equation}
    \dot{n}(\mathbf{\Theta})=\frac{N_{\rm{D}}(\mathbf{\Theta})}{\mathcal{D(\mathbf{\Theta})}}\frac{1+z}{T}/\frac{dV_c}{dz},\label{eqn:inverse}
\end{equation}
and $N_{\rm{D}}(\mathbf{\Theta})$ can be inferred from the observed catalogue.

Marginalising over $m_1$ and $m_2$ in equation (\ref{eqn:inverse}) on both sides, we have:
\begin{equation}
    \mathcal{R}(z)=I(z)|_{w_i=1/\mathcal{D}_i}\frac{1+z}{T}/\frac{dV_c}{dz},
    \label{eqn:Rkde}
\end{equation}

where $I(z)$ is the 1-D intensity function (number density) over $z$, of the observed catalogue, with weights $w_i$ equal to the corresponding $1/\mathcal{D}(\Theta_i)$. The intensity can be calculated with an empirical density estimator of the (synthetic) catalogue or with the (non-normalized) weighted Kernel Density Estimator (KDE, see appendix \ref{app:KDE}). By simulating a number of different realisations of the Universe population inferred from $I(z)$, the confidence intervals of the estimated $I(z)$ are obtained. 

Here, in order to include the observational uncertainties and give at the same time an estimate of the Confidence Intervals of the KDE, a specific bootstrap is employed, as follows:
\begin{enumerate}
    \item The \GWT returns a catalogue $\{\mathbf{\Theta}\}_{\rm{true}}$, where $\mathbf{\Theta}^k_{\rm{true}}$ denotes the true parameters of the $k$-th events in the catalogue, including its $m_1$, $m_2$, $z$ and their uncertainties $\delta m_1$, $\delta m_2$, $\delta z$. $k$ goes from 1 to $N$, where $N$ is the total number of events.  
    \item Shift the true values $\mathbf{\Theta}^k_{\rm{true}}$ according to a multivariate Gaussian centred on the true values and with covariance matrix given by the corresponding uncertainties. Therefore, a new catalogue $\{\mathcal{\mathbf{\Theta}}^k\}_{\rm{obs}}$ is obtained. 
    \item Estimate the detection probabilities $\mathcal{D}( \mathbf{\Theta}^k_{\rm{obs}})$.
    \item Calculate the (unnormalized) weighted KDE of $\{\mathcal{\mathbf{\Theta}}^k_{\rm{obs}}\}$, with weights $1/\mathcal{D}(\mathbf{\Theta}^k_{\rm{obs}})$. The intensity is estimated on the log and then transformed back to the original support; this ensures that there isn't any leakage of the intensity on negative support.
    \item Take $M\equiv\sum_{i=k}^{N}1/\mathcal{D}( \mathbf{\Theta}^k_{\rm{obs}})$, as an estimate of the total number of mergers in the Universe in the parameters space. 
    \item Generate a realization $\widetilde{M}$ according to a Poisson with expected value $M$. 
    \item Draw a sample of size $\widetilde{M}$ from the KDE calculated in step 4. This is the equivalent of generating a new Universe sample, based on the estimate of the distribution in the Universe from the observations.
    \item For each of the events in the above sample, calculate its detection probability. Draw $\widetilde{M}$ uniform random values,  $U$, in between 0 and 1, and drop the observations for whose detectability is less than $U$. A smaller sample is left, which represents a new realization of the detected catalogue. 
    \item Estimate the covariance for each observation of the generated data set $\mathbf{\Sigma_{{\Theta}^k}}$ using interpolation of the original catalogue, and shift them according to a multivariate Gaussian centred on each observation and with covariance matrix $\mathbf{\Sigma_{{\Theta}^k}}$. 
    \item Estimate their detection probability.
    \item  Calculate the (unnormalized) weighted KDE of the above mentioned sample. 
    \item Repeat steps from 6 to 11 for a large number of times $B$ (we find $B$=200 is large enough). Calculate the confidence intervals according to the assigned quantiles.
    \item Based on the results above, we can correct for the bias introduced by the measurement uncertainties. We estimate the bias between the bootstrap intensities and the intensity estimated on the observed sample, and use it to rescale the observed sample's estimated intensity.
    
\end{enumerate}

\begin{figure}
    \centering
    \includegraphics[width=0.5\textwidth]{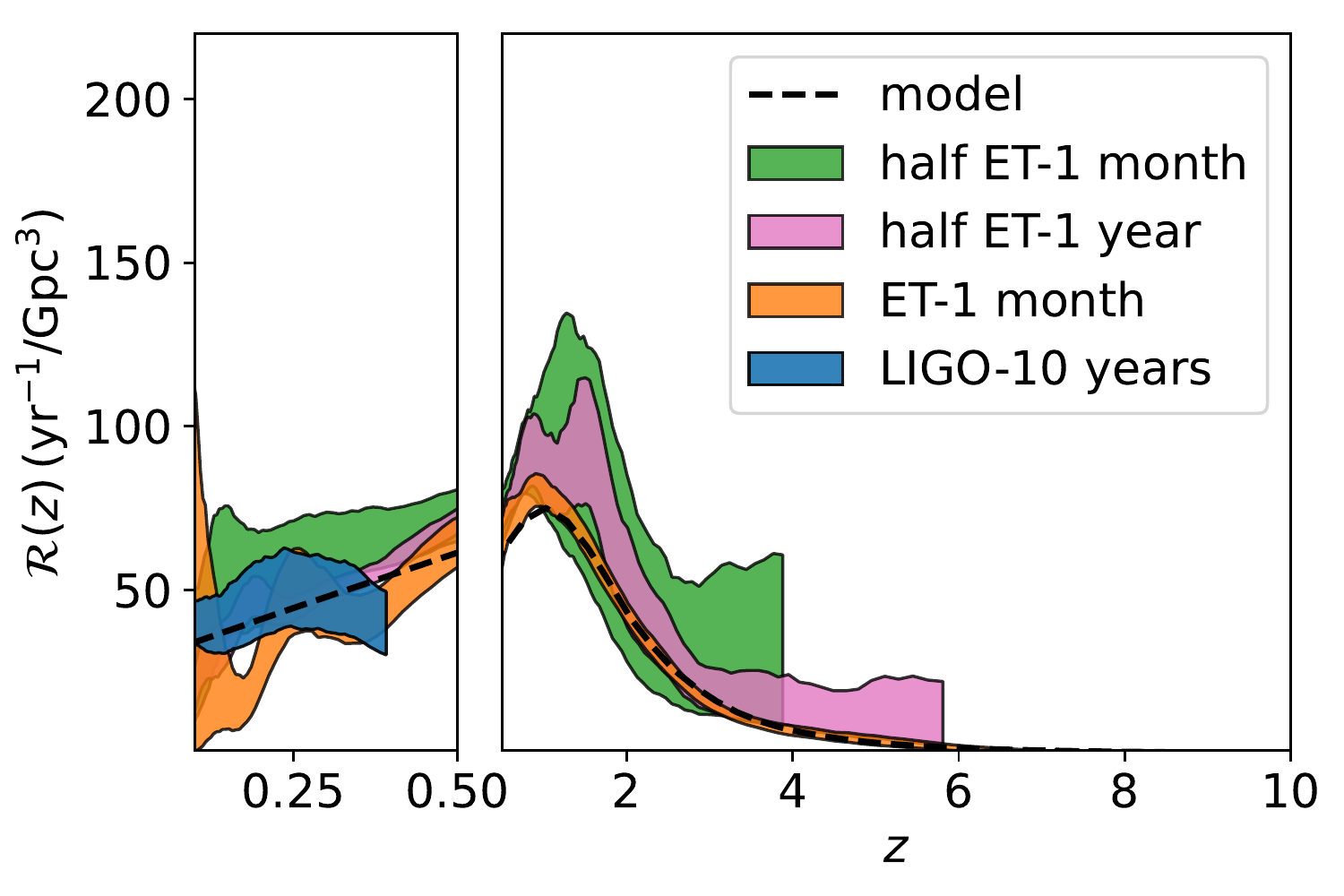}
    \caption{The merger rate as function of the redshift, reconstructed with equation (\ref{eqn:pm1kde}) from the synthetic aLIGO, ET and half-ET catalogues.}
    \label{fig:Rzkde}
\end{figure}

Similarly, the merger rate in the Universe as function of the mass of primary BH is:
\begin{equation}
    \mathcal{R}(m_1)=p(m_1)\times \dot{n}_{\rm{tot}}=I(m_1)|_{w_i=1/\mathcal{D}_i}/T,
    \label{eqn:pm1kde}
\end{equation}
where $p(m_1)$ is the normalized mass function of the primary BH, $\dot{n}_{\rm{tot}}$ is the total number of BBH mergers in the Universe in a unit local time. $I(m_1)$ is the 1-D intensity function of $m_1$, calculated from the observed catalogue. The procedures of obtaining $I(m_1)$ is the same as that in the calculation of $I(z)$.   
In figures \ref{fig:Rzkde} and \ref{fig:Pm1kde} we plot the $\mathcal{R}(z)$ and $\mathcal{R}(m_1)$ reconstructed with equations \ref{eqn:Rkde} and \ref{eqn:pm1kde}  from the synthetic aLIGO, ET and half-ET catalogues. The shaded areas are the $95\%$ confidence intervals. 

\begin{figure}
    \centering
    \includegraphics[width=0.5\textwidth]{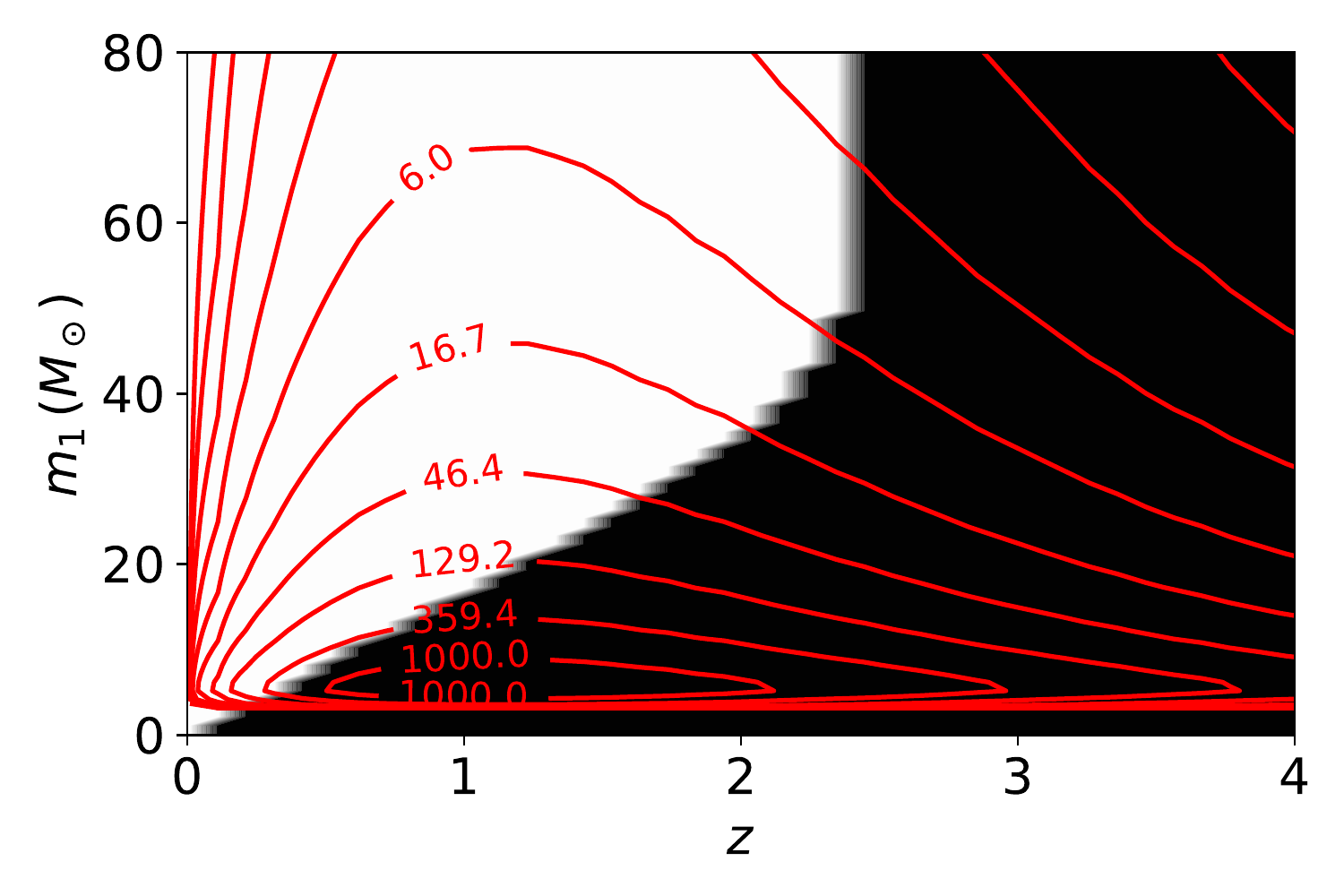}
    \caption{Two-dimensional theoretical merger rate as function of redshift and primary masses. The black region marks the region in the parameters space that is not reachable by aLIGO (single aLIGO detector at its nominal sensitivity).}
    \label{fig:2Dtheory}
\end{figure}

\begin{figure}
    \centering
    \includegraphics[width=0.5\textwidth]{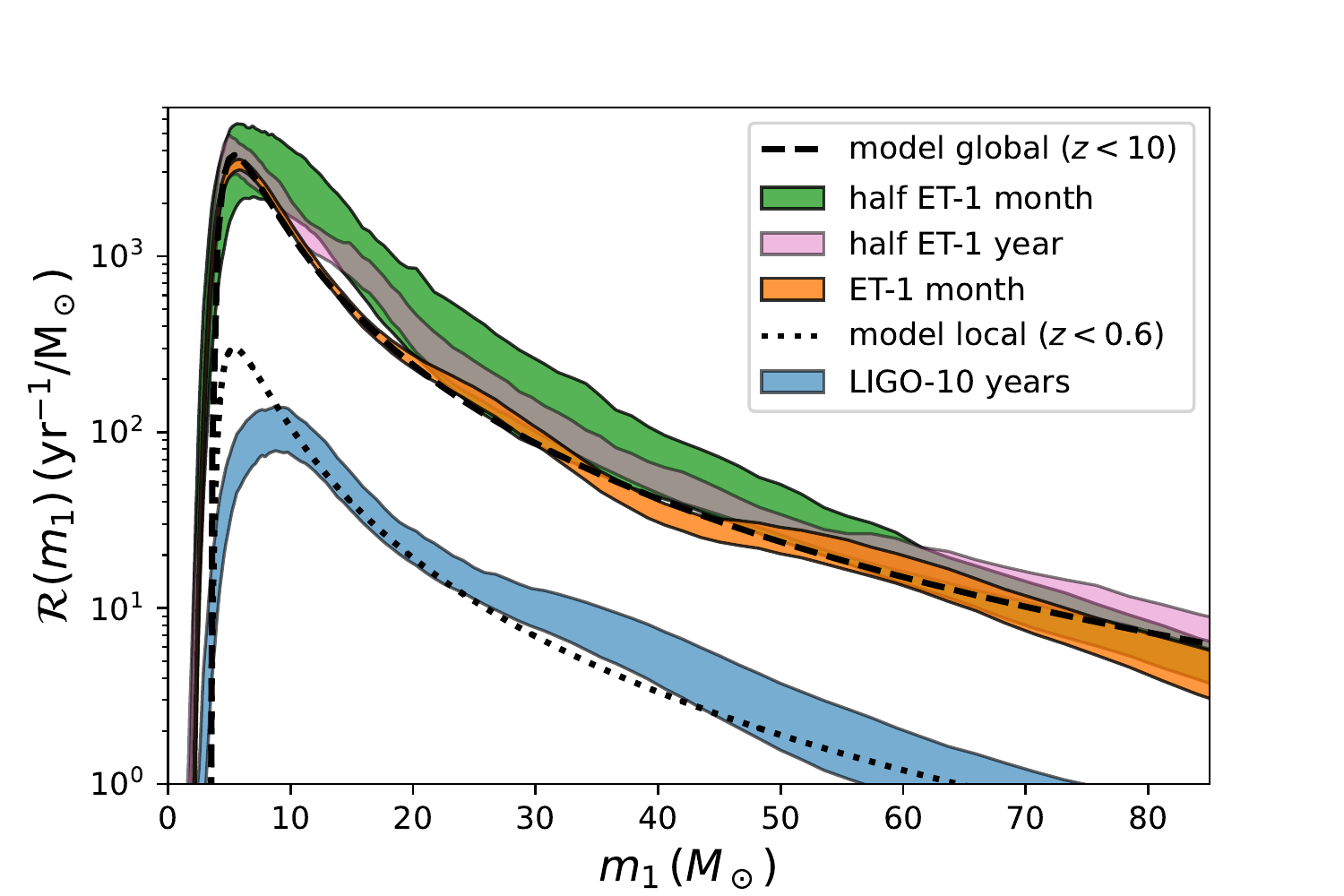}
    \caption{The merger rate as function of the primary BH mass, reconstructed with equation (\ref{eqn:pm1kde}) from the synthetic aLIGO, ET and half-ET catalogues.}
    \label{fig:Pm1kde}
\end{figure}

We note in figure \ref{fig:Rzkde} that, for aLIGO's observation, the obtained $\mathcal{R}(z)$ in $z>0.3$ is an underestimate of the underlying model. The reason is that aLIGO's detectable range of BBH masses is increasingly incomplete towards higher redshift. An intuitive illustration is in Figure \ref{fig:2Dtheory}. There, we plot the two dimensional merger rate density as function of $m_1$ and $z$. The underlying black region marks the parameters space region where the detectability is essentially zero, i.e., the region beyond aLIGO's horizon. When plotting Figure \ref{fig:2Dtheory}, we set the cut at $\mathcal{D}(z,m_1,m_2=m_1)<10^{-2}$. We can see the fraction of the black region in the mass spectrum is increasing towards high redshift. This corresponds to the fact that BBH with small masses cannot be detected with aLIGO at higher redshift.

Similarly, in figure \ref{fig:Pm1kde}, we see $\mathcal{R}(m_1)$ is deviating the theory towards low $m_1$. To compare the estimation with the model, we limit both the aLIGO catalogue and the theory in the range $z<0.6$. Therefore, the meaning of $\mathcal{R}(m_1)$ is no longer the merger rate density of the complete Universe, but that of a local Universe up to 0.6, which is calculated with:
\begin{equation}
    \mathcal{R}(m_1)|_{z<0.6}=p(m_1)\int^{0.6}_0\mathcal{R}(z)\frac{dV}{dz}/(1+z)dz, 
\end{equation}
where $p(m_1)$ is the normalized mass function. 

It is unlike the results with the Bayesian inference method, where constraints can still be placed in regimes with no detection. The constraints on $\mathcal{R}(z)$ and the mass function are also less tight for the nonparametric method comparing with the parametric one. The difference is rooted in the degree of prior knowledge we assumed with the population model in both methods: In the parametric method, we assume the function forms which are exactly the same as the underlying population model that was used to generate simulated catalogue. It represents a too optimistic extreme. For in reality, we would never be able to guess a perfect function representation of the underlying population. On the other hand, in the nonparametric method, we do not impose any assumption on the properties of the population model. It represents a conservative extreme. For in reality, we more or less know the general shape or trends of the population. 


\section{Conclusion and Discussion}\label{sec:candd}
In this paper, we describe the details of how the \GWT simulates observations of BBH mergers with different ground-based GW detectors. As a demonstration of its scientific application, we use the synthetic catalogues to reconstruct and constrain the cosmic merger rate and mass function of the BBH population. The comparison amongst the results from one month (half) ET observation and ten years aLIGO observations shows the overwhelming advantages of the 3rd generation detectors over the 2nd generation, especially at a redshift higher than $\sim2$, around where the cosmic merger rate is believed to peak. 

The reconstruction and constraining on $\mathcal{R}(z)$ and $p(m_1)$ are performed with two methods, namely, 1) a Bayesian method, assuming a parameteric formula of $\mathcal{R}(z)$ and $p(m_1)$ and 2) a KDE method with non-parameteric $\mathcal{R}(z)$ and $p(m_1)$. For the catalogue of ET observations, the results from both methods are qualitatively the same. However, for aLIGO, the Bayesian method can put constraints on $\mathcal{R}(z)$ beyond the detecting limit, where the non-parameteric method lost its ability completely. The $m_1$ dependence is also much more accurately reconstructed by the Bayesian method than the non-parameteric method for aLIGO's catalogue. The difference is due to the extra information placed by assuming a specific parameterisation of the population model in the Bayesian method. 
In our example, the underlying $\mathcal{R}(z)$ and $p(m_1)$ used to generate the catalogues are the same with those assumed in the Bayesian method; while in the KDE method, no assumption of the general shape of $\mathcal{R}(z)$ and $f(m_1)$ are placed, not even the assumption of its smoothness. These represent the two extreme situations in the differential merger rate reconstruction.  

In reality, the underlying shape of $\mathcal{R}(z)$ and $f(m_1)$ are unknown and they could be very different with what we used. The parameterised Bayesian method may give biased inference and underestimated uncertainties. Furthermore, any unexpected structures in $\mathcal{R}(z)$ and $f(m_1)$ cannot be recovered with the parameterised Bayesian method. For instance, there can be additional peaks in the BH spectrum corresponding to pulsational pair-instability supernovae and primordial BHs. On the other hand, although the function form of the merger rate is unknown, the general trends should be more or less known. A better reconstruction method should lie in between these two extreme methods. 

An alternative method is to parameterise $\mathcal{R}(z)$ and $p(m_1)$ as piecewise constant functions and use the Bayesian inference to obtain the posterior of the values in each bin, as in \cite{2019ApJ...886L...1V}. It has the advantage that no assumptions on the shapes of $\mathcal{R}(z)$ and $p(m_1)$ is needed, like the Kernel density estimator method. The disadvantage of this method is that MCMC sampling in high dimension needs to be performed. If one assumes that the $p(m_1)$ is independent of $z$, there are more than $N+M$ free parameters, where $N$ and $M$ are the number of bins in $z$ and $m_1$ respectively. If we want the resolutions to be $0.5$ in $z$ and $1\,M_\odot$ in $m_1$, then the dimension of the parameter space is $>80$ for ET catalogue. In the more general case, when one includes the dependence of $p(m_1)$ on $z$, the number of parameters becomes $N\times M$, therefore the dimension of the parameter space goes catastrophically to $>1200$. 

Another conclusion we draw about the early phase ET (half-ET) is that, half-ET can also give a good constraint on $\mathcal{R}(z)$ at high redshift Universe. However, both $\mathcal{R}(z)$ and primary mass function are less accurately constrained compared with the full ET, even using a larger number of detected events from a longer observation duration. The reason is that the full ET determines smaller uncertainties on the physical parameters of the GW events.

In this paper, we compared the results from a single aLIGO detector with that of ET. In reality, there will always be multiple detectors working as a network (e.g. LIGO-Virgo-KAGRA), which is expected to have better results than that a single aLIGO. However, we do not expect a qualitatively different conclusion from that ET (as a representative of the 3rd generation GW detectors) has an overwhelming ability in constraining the BBH population properties over the 2nd generation detectors, even in a form of a network. We also anticipate that the same conclusion applies to other 3rd generation GW  detectors, \textit{e.g., }the Cosmic Explorer.

\appendix 
\section{Kernel density estimation}\label{app:KDE}
Kernel density estimation (KDE) is one of the most famous method for density estimation. For the univariate case is defined as:
\begin{equation}
    \hat{f}(x,h)= \frac{1}{n h}\sum_{i=1}^{n} K (\frac{x-X_i}{h}) ,
\end{equation}

\noindent
where $K$ is the chosen kernel and $h$ is the so-called bandwidth parameter. In the case of a Gaussian kernel, $h$ can be thought as the standard deviation of the normal density centred on each data point. 
The choice of the kernel doesn't really affect the density estimate, if the number of observations is high enough; what does instead affect it is the choice of the bandwidth parameter $h$, that has to be estimated with one of the available methods. The most common methods are rule-of-thumb like Scott’s rule (\cite{Scott1992}) $H = n^{
-2/(m+4)} \Sigma$ and Silverman’s rule (\cite{Silverman1986}) $H = \frac{4} {n(m + 2)}^{2/(m+4)} \Sigma$, where $m$ is the number of the variables, n is the sample size and $\Sigma$ is the empirical covariance matrix. A more accurate but computationally expensive method is Cross-validation bandwidth selection described in \cite{Scott1987}.

As introduced in \cite{RePEc:ecm:emetrp:v:64:y:1996:i:5:p:1001-44}, KDE can be adapted to the case where each observation is attached to a weight that reflects its contribution to the sample. The weighted kernel density estimator can be written as:
\begin{equation}
    \hat{f}(x,h)= \frac{1}{n h}\sum_{i=1}^{n} w_i K (\frac{x-X_i}{h}) ,
\end{equation}
to ensure that the result of the density estimation has still the properties of a density, it is necessary to normalize the weights so that $\sum_{i=1}^{n} w_i = 1$.

KDE is easily extendable to the multivariate case, and used to estimate densities $f$ in $R^p$ based on the same principle: perform an average of densities centred at the data points.
For the multivariate case, kernel density estimation is defined as:
\begin{equation}
   \hat{f}(x,H)= \frac{1}{n |H|^{\frac{1}{2}}} \sum_{i=1}^{n} K ( H^{-\frac{1}{2}}(x-X_i)) ,
\end{equation}

\noindent
where $K$ is the multivariate kernel, a p-variate density that depends on the bandwidth matrix $H$, a $p\times p$ symmetric and positive definite matrix.

\end{document}